\documentclass{elsart}

\usepackage[dvips]{graphicx}

\begin{document}

\begin{frontmatter}

\title{Transverse Instability of Solitons Propagating on Current-Carrying Metal Thin Films}

\author{R. Mark Bradley}

\address{
Department of Physics,
Colorado State University,
Fort Collins, CO 80523 USA
}

\begin{abstract}
Small amplitude, long waves travelling over the surface of a 
current-carrying metal thin film are studied.  The equation of motion
for the metal surface is determined in the limit of high applied currents, when 
surface electromigration is the predominant cause of adatom motion.  If the surface
height $h$ is independent of the transverse coordinate $y$, the equation of motion
reduces to the Korteweg-de Vries equation. 
One-dimensional solitons (i.e., those
with $h$ independent of $y$) are shown to be unstable against perturbations
to their shape with small transverse wavevector.
\end{abstract}

\end{frontmatter}

\section{Introduction}

The motion of the free surface of an incompressible fluid
has been the subject of much research, from the 19th century
to the present day.  Of the many fascinating phenomena that have been 
investigated, 
soliton propagation is certainly among the most intriguing \cite{Drazin,Ablowitz}.

Solitons were first observed in 1834 by John Scott Russell on the 
surface of a narrow, water-filled canal  \cite{Russell}.  
It was not until 1895, however, that
it was demonstrated that solitons are solutions to the Korteweg-de Vries
(KdV) equation, the equation of motion for small amplitude, 
long gravity waves \cite{KdV}.

Although it is not yet widely appreciated, there are nontrivial
{\it electrical} free boundary problems. 
When an electrical current passes through a piece of solid metal, 
collisions between the conduction electrons and the metal atoms at the
surface lead to drift of these atoms.  This phenomenon, which is 
known as surface electromigration (SEM), can cause a solid metal surface
to move and deform 
\cite{Krug_and_Dobbs,Kraftapl,Suojap,Mohanjap,Krug1,Krug2,Gengor1,Gengor2,Gengor3,Mohanepl,Mohanpre}.  
The free surface of a metal film moves in response
to the electrical current flowing through the bulk of the film, in much
the same way that flow in the bulk of a fluid affects the motion of 
its surface.  However, the analogy is not perfect --- the 
boundary conditions are very different in the two problems.

Just as solitary waves can propagate over the surface of
a shallow body of water, solitons can propagate over the
free surface of a current-carrying metal thin film \cite{Krug1,Gengor2,Gengor3,Bradley}.  Let 
the applied electric field be in the $x$ direction and take the 
$z$ axis to be normal to the undisturbed metal surface.  If the surface
height $h$ is independent of $y$, the equation of motion for small amplitude, long waves on the
free surface is the KdV equation
in the limit of high applied currents \cite{Bradley}.  The \lq\lq one-dimensional" 
solitons satisfying the KdV equation are protrusions that propagate
in the direction of the applied electric field.  
Their velocity decreases linearly with amplitude;  
in contrast, the velocity of a soliton on the free surface of a body of water
increases with amplitude. 

A soliton propagating along a narrow, shallow channel of water is stable
against small perturbations to its shape \cite{Jeffery,Benjamin}.  
The issue of whether or not a one-dimensional (1D) soliton
is stable is more complex if it is travelling over the surface of a sheet of water.
The equation of motion for weakly two-dimensional,
small amplitude, long waves on a fluid sheet
is the Kadomtsev-Petviashvili (KP) equation 
\begin{equation}
(u_t+6uu_x+u_{xxx})_x+3\sigma^2u_{yy}=0,
\label{KP}
\end{equation}
where $u_x\equiv\partial u/\partial x$ and so forth \cite{Ablowitz}.  The KP equation reduces
to the KdV equation if $u$ is independent of $y$.
In the regime in which the effects of gravity dominate those of surface tension,
$\sigma^2=+1$ and the equation of motion is known as the KPII equation.  
1D solitons are stable in this case \cite{KP}.  The equation of motion in the
regime in which surface tension is dominant is Eq. (\ref{KP}) with $\sigma^2=-1$,
i.e., the KPI equation.  In this case, the 1D solitons are unstable \cite{KP}.

The KP equation is not the only possible generalization of the KdV equation 
to two dimensions.  The Zakharov-Kuznetsov (ZK) equation
\begin{equation}
u_t+(u_{xx}+u_{yy})_x+uu_x=0
\label{ZK}
\end{equation}
is another; it describes deviations from the average
ion density in an ion-acoustic plasma permeated 
by a strong uniform magnetic field \cite{ZK}.
In this case, 1D solitons are unstable against perturbations of sufficiently long
transverse wavelength \cite{LS,Infeld1,Infeld2,FI,AR}.

To what extent does electromigration-induced soliton propagation resemble
soliton propagation over the free surface of a fluid?  Consider a long, rectangular,
current-carrying ribbon of metal.  Suppose that the metal ribbon has small thickness,
and that it is passivated along all but one edge (Fig. 1).  Solitons
travelling along the free surface of the metal film satisfy the KdV equation and are stable \cite{Bradley,Jeffery,Benjamin} --- 
this is the analog of a soliton moving along a narrow channel of water.

\vskip 20pt

\includegraphics[scale=0.75]{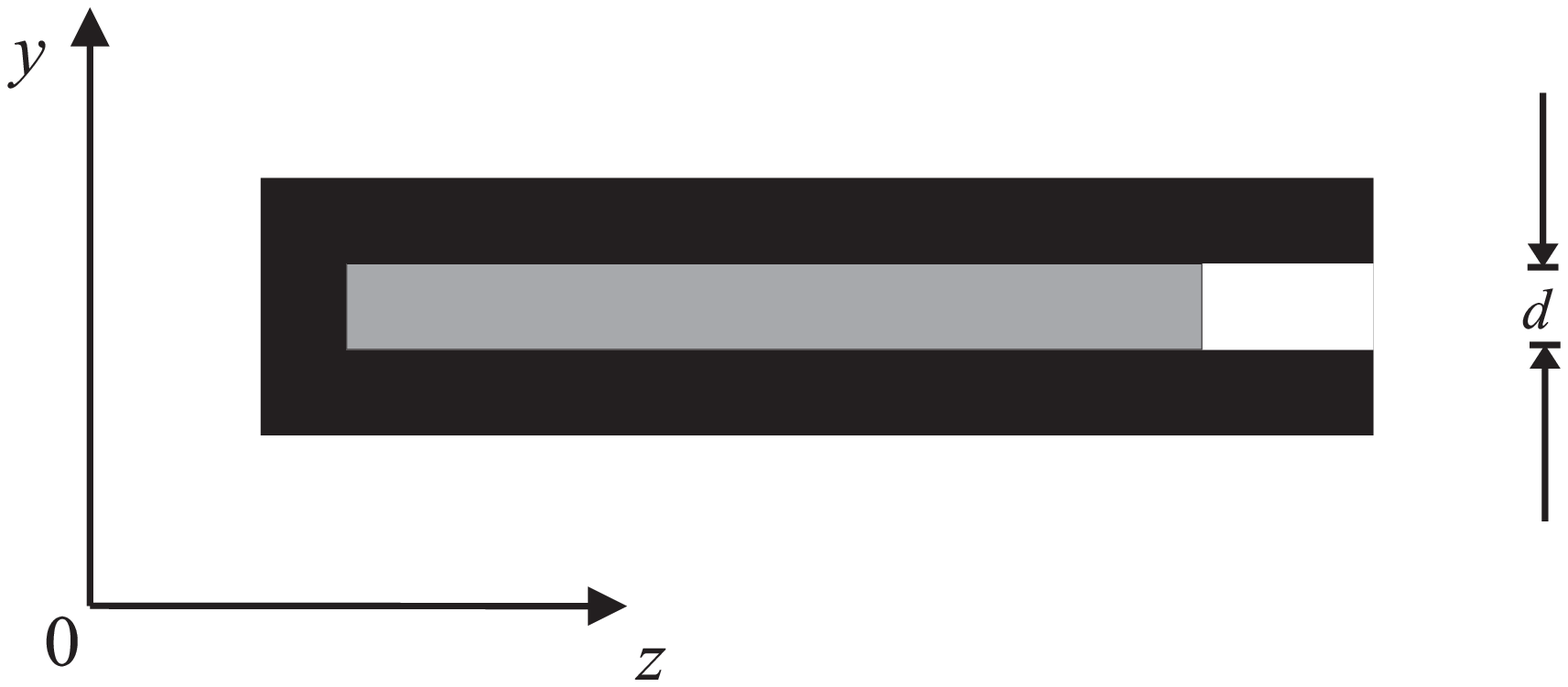}

\vskip 20pt

\centerline{
\vbox{\hsize=4.5 in
{\bf Figure 1.} Cross section of the metal ribbon. 
The passivating layer (shown in black) is an electrical insulator and 
covers all but one edge of the ribbon (shown in gray). 
Since the thickness $d$ of the ribbon is small, the location
of the free surface of the metal is to a good approximation
independent of $y$.  The electric field points in the $x$ direction
if the free surface is undisturbed.
}
}

\vskip 20pt

As we have seen, a 1D soliton travelling over the surface of a fluid sheet
may or may not be stable, depending on the value of the surface tension.
We now pose the analogous question for electromigration:
Is a 1D soliton travelling over the surface of a metal thin film
of infinite extent stable?  Answering this question is the primary goal of this paper.  

To begin, we shall determine the equation of motion for small amplitude, long waves
on the surface of a current-carrying metal film for the general case
in which the surface height $h$ depends on both $x$ and $y$.  This equation differs
from both the KP and ZK equations, and seems not to have been studied previously.
We then demonstrate that the one-dimensional solitons are unstable against 
perturbations with small transverse wavevector.

\vfill\eject

\section{Equations of motion}

Consider a metal film of thickness $h_0$ deposited on the plane surface of
an insulating substrate.  We take the $z$ axis to be normal to the substrate 
surface and locate the origin in this plane.  A constant current flows
through the film in the $x$ direction, and the electric field within 
the metal is $E_0\hat x$.

Now suppose that the upper surface of the 
film is perturbed (Fig. 2). Let the 
outward-pointing unit normal to this surface be $\hat n$. 
The height of the film's surface above the substrate $h$
depends on $x$, $y$ and $t$.  The upper film surface will evolve in the 
course of time due to the effects of SEM and surface self-diffusion.  We 
assume that the current flowing through the film is held fixed.

\vskip 20pt

\includegraphics[scale=0.85]{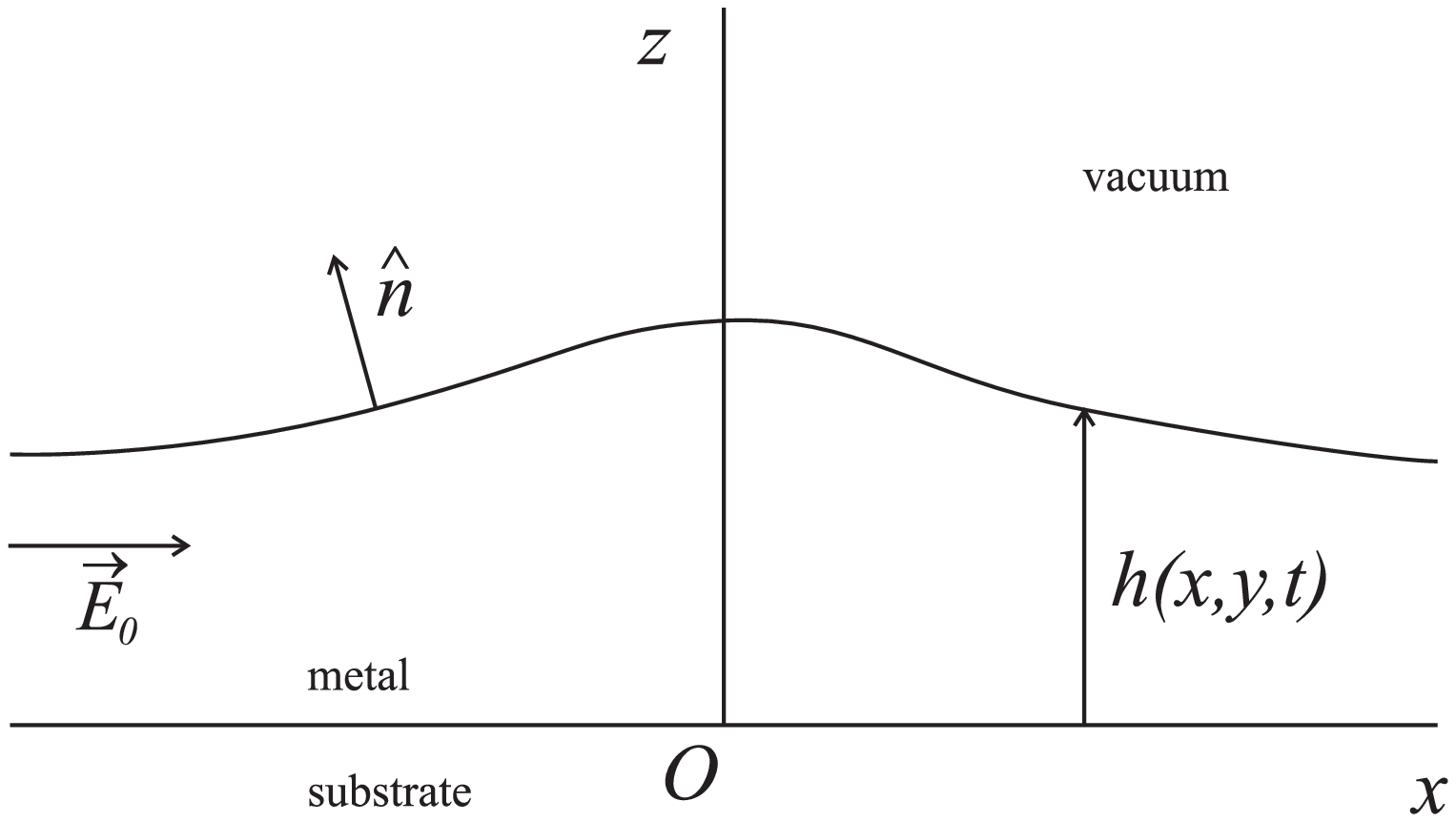}

\vskip 20pt

\centerline{
\vbox{\hsize=4.5 in
{\bf Figure 2.} The current-carrying metal thin film.  The height of the free surface 
above the substrate, $h$, depends on $x$, $y$ and $t$.  The outward-pointing
unit normal to the free surface is $\hat n$, and the electric field far from the 
perturbation is $\vec E_0$.
}
}

\vskip 20pt

The electrical potential $\Phi=\Phi(x,y,z,t)$ satisfies
the Laplace equation
\begin{equation}
\nabla^2\Phi=0
\label{orig_Laplace}
\end{equation}
and is subject to the boundary condition  
$\hat n\cdot\vec\nabla\Phi = 0$ on the upper surface and 
$\hat z\cdot\vec\nabla\Phi = 0$ on the lower.
More explicitly, we have
\begin{equation}
\Phi_z(x,y,h,t) =h_x\Phi_x(x,y,h,t) + h_y\Phi_y(x,y,h,t)
\label{orig_topbc}
\end{equation}
and
\begin{equation}
\Phi_z(x,y,0,t)=0.
\label{orig_bottombc}
\end{equation}  
If the initial perturbation
is localized in the $x$ direction, we will also have
\begin{equation}
h(x,y,t)\to h_0
\label{orig_hbc}
\end{equation}
for $x\to\pm\infty$ and, furthermore,
\begin{equation}
\Phi_x(x,y,z,t)\to -E_0,
\label{orig_sidebc1}
\end{equation}
and
\begin{equation}
\Phi_y(x,y,z,t)\to 0
\label{orig_sidebc2}
\end{equation}
for $x\to\pm\infty$, arbitrary $y$, and $0\le z \le h_0$.

We assume that the mobility of the metal atoms is negligible at the 
metal-insulator interface, so that the form of that interface remains
planar for all time.  Further, in the interest of simplicity, we assume
that the applied current is high enough that the effects of SEM are 
much more important than those of capillarity.  The surface atomic
current $\vec J_a$ is then proportional to the electrical current at the surface.
Explicitly, $\vec J_a = -\nu Mq\vec\nabla\Phi$, where the $\nu$
is the areal density of the mobile surface atoms, $q$ is their effective 
charge, and $M$, the adatom mobility, will be assumed to have negligible anisotropy.
When there is a net influx of atoms into a small surface element, the surface
element will move.  The normal velocity of the metal surface is 
\begin{equation}
v_n = qM\nu\Omega \nabla^2_S \Phi,
\end{equation}
where $\Omega$ is the atomic volume and $\nabla^2_S$ is the surface (Beltrami)
Laplacian \cite{Weatherburn}.  Note that $h_t = \sqrt{g} v_n$, where $g\equiv 1+h_x^2+h_y^2$
is the determinant of the metric tensor for the surface.  Using this relation
and writing out the surface Laplacian explicitly, we have
\begin{eqnarray}
{1\over{qM\nu\Omega}}{{\partial h}\over{\partial t}}
&=& \partial_x\left\{ {1\over{\sqrt{g}}}
[(1+h_y^2)\partial_x - h_xh_y\partial_y] \Phi(x,y,h, t) \right\}\nonumber\\
& &+
\partial_y\left\{ {1\over{\sqrt{g}}}
[(1+h_x^2)\partial_y - h_xh_y\partial_x] \Phi(x,y,h, t) \right\}.\label{orig_eom}
\end{eqnarray}
Here $\partial_x$ is the total derivative with respect to $x$, and $\partial_y$
is similarly defined.  (Since $h$ depends on $x$, we have 
$\partial_x\Phi(x,y,h,t) = \Phi_x(x,y,h,t)+h_x\Phi_z(x,y,h,t)$, for example.)
Together, Eqs. (\ref{orig_Laplace}) - (\ref{orig_eom}) completely describe the 
nonlinear dynamics of the film surface.

\section{Asymptotic analysis}

We wish to study the propagation of a localized disturbance whose 
amplitude is small compared to $h_0$, and whose width is much larger
than $h_0$.  To do so, we will use multiple scale asymptotic
analysis \cite{Drazin}.  We put $h=h_0 + a\zeta$, where the constant $a$ is a
measure of the amplitude of the disturbance and $\zeta=\zeta(x,y,t)$
is of order unity.  Let $l$ be the characteristic width of the 
initial disturbance.  We will study the limit in which both
$\alpha\equiv a/h_0$ and $\delta\equiv h_0/l$ are small.
More precisely, we shall consider the limit in which $\alpha$
and $\delta$ tend to zero, but $\alpha/\delta^2$ remains finite.
It is in this limit that the effects of nonlinearity and dispersion
balance, and solitons can propagate.

To simplify the description of the problem, we will set $\Phi = -E_0 x +\phi$,
so that $\vec\nabla\phi\to 0$ for $x\to\pm\infty$.  We also introduce 
the dimensionless quantities
$\tilde x\equiv x/l$, $\tilde y\equiv y/l$, $\tilde z\equiv z/h_0$,
$\tilde t\equiv (\vert q\vert ME_0 \nu\Omega  /h_0 l)t$,
and
$\tilde \phi(\tilde x, \tilde y, \tilde z, \tilde t)\equiv \phi(x,y,z,t)/(E_0 l)$.
Eq. (\ref{orig_Laplace}) becomes
\begin{equation}
\tilde\phi_{\tilde z\tilde z} + \delta^2(\tilde\phi_{\tilde x\tilde x} + \tilde\phi_{\tilde y\tilde y})   = 0,
\label{scaled_Laplace}
\end{equation}
which applies for $0\le\tilde z\le 1+\alpha\zeta(\tilde x,\tilde y, \tilde t)$ and all $\tilde x$ and $\tilde y$.
The boundary conditions are 
\begin{equation}
\tilde\phi_{\tilde z} = 0 \qquad {\rm for} \,\,\tilde z = 0,
\label{scaled_bottombc}
\end{equation}
\begin{equation} 
\tilde\phi_{\tilde z} = \alpha\delta^2(\tilde\phi_{\tilde x} - 1)\zeta_{\tilde x} 
+ \alpha\delta^2\tilde\phi_{\tilde y}\zeta_{\tilde y}
\qquad {\rm for} \,\,\tilde z =1+\alpha\zeta(\tilde x,\tilde y, \tilde t),
\label{scaled_topbc}
\end{equation}
\begin{equation}
\zeta\to 0 \qquad {\rm for} \,\,\tilde x\to\pm\infty ,
\label{zeta_bc}
\end{equation}
and
\begin{equation}
\tilde\phi_{\tilde x}\hat x + \tilde\phi_{\tilde y}\hat y \to 0 \qquad {\rm for} \,\, 0\le \tilde z\le 1
\,\, {\rm and}\,\, \tilde x\to \pm\infty .
\label{scaled_sidebc}
\end{equation}
Finally, the equation of motion for the free surface of the film is 
\begin{eqnarray}
\sigma_q\alpha\zeta_{\tilde t} &=&
\partial_{\tilde x} 
\left\{ {1\over{\sqrt{g}}}\left [ 
(1+\alpha^2\delta^2\zeta^2_{\tilde y})(\partial_{\tilde x}\tilde\phi -1)
- \alpha^2\delta^2\zeta_{\tilde x}\zeta_{\tilde y} \partial_{\tilde y}\tilde\phi\right ]\right\}\nonumber\\
& &+
\partial_{\tilde y}
\left\{ {1\over{\sqrt{g}}}\left [ 
(1+\alpha^2\delta^2\zeta^2_{\tilde x})\partial_{\tilde y}\tilde\phi
- \alpha^2\delta^2\zeta_{\tilde x}\zeta_{\tilde y} (\partial_{\tilde x}\tilde\phi-1)\right ]\right\},
\label{scaled_eom}
\end{eqnarray}
where $\tilde z = 1+\alpha\zeta(\tilde x, \tilde y, \tilde t)$ and $\sigma_q\equiv q/\vert q\vert$.
Note that $g = 1 + \alpha^2\delta^2 (\zeta^2_{\tilde x} + \zeta^2_{\tilde y})$.

We next introduce the scaled variables
\begin{equation}
\xi\equiv {{\alpha^{1/2}}\over\delta} (\tilde x + \sigma_q\tilde t ),
\end{equation}
\begin{equation}
\eta\equiv {{\alpha^{1/2}}\over\delta} \tilde y ,
\end{equation}
\begin{equation}
\tau \equiv {{\alpha^{3/2}}\over\delta}\tilde t ,
\end{equation}
and
\begin{equation}
\psi\equiv {{\alpha^{1/2}}\over\delta}\tilde\phi .
\end{equation}
$(\xi ,\eta , \tau)$ is a moving coordinate system that translates
with velocity $v_0\equiv -qME_0\nu\Omega/h_0$ relative to the laboratory frame. 
$v_0$ is the velocity of surface waves in the limit of vanishing amplitude
and wavevector \cite{Krug1}.

For convenience, we drop the tilde on $z$.  Laplace's equation becomes
\begin{equation}
\psi_{zz} + \alpha(\psi_{\xi\xi} + \psi_{\eta\eta})  = 0,
\label{Laplace}
\end{equation}
and this applies for $0\le z \le 1+\alpha\zeta(\xi,\eta, \tau)$ and all $\xi$ and $\eta$.
In terms of the scaled variables, 
Eqs. (\ref{scaled_bottombc}) - (\ref{scaled_eom}) are
\begin{equation}
\psi_z = 0 \qquad {\rm for} \,\, z = 0,
\label{bottombc}
\end{equation}
\begin{equation} 
\psi_z = \alpha^2[(\psi_\xi - 1)\zeta_\xi + \zeta_\eta\psi_\eta ] \qquad {\rm for} \,\, z 
= 1+\alpha\zeta(\xi,\eta, \tau) ,
\label{topbc}
\end{equation}
\begin{equation}
\zeta\to 0 \qquad {\rm for} \,\,\xi\to\pm\infty ,
\label{new_zeta_bc}
\end{equation}
\begin{equation}
\psi_\xi\hat x +\psi_\eta\hat y\to 0 
\qquad {\rm for} \,\, 0\le z\le 1\,\,{\rm and}\,\,\xi\to \pm\infty ,
\label{sidebc}
\end{equation}
and
\begin{eqnarray}
\alpha\zeta_\xi + \sigma_q\alpha^2\zeta_{\tau} &=&
\partial_\xi
\left\{ {1\over{\sqrt{g}}}\left [ 
(1+\alpha^3\zeta^2_{\eta})(\partial_\xi\psi -1)
- \alpha^3\zeta_{\xi}\zeta_{\eta} \partial_\eta\psi\right ]\right\}\nonumber\\
& &+
\partial_\eta
\left\{ {1\over{\sqrt{g}}}\left [ 
(1+\alpha^3\zeta^2_{\xi})\partial_\eta\psi
- \alpha^3\zeta_{\xi}\zeta_{\eta} (\partial_\xi\psi-1)\right ]\right\}
\label{eom}
\end{eqnarray}
for $z = 1+\alpha\zeta(\xi,\eta, \tau)$.  In terms of the scaled variables,
$g = 1 + \alpha^3(\zeta_\xi^2+\zeta_\eta^2)$.
One advantage of introducing the scaled variables is now manifest: 
$\delta$ does not appear explicitly in Eqs. (\ref{Laplace}) - (\ref{eom}).

We shall now begin our analysis of the small $\alpha$ limit.  We assume that 
for small $\alpha$ and fixed $\xi$ and $\tau$, there is a solution with
\begin{equation}
\zeta = \sum_{n=0}^\infty \alpha^n\zeta_n(\xi,\eta, \tau)
\label{zeta_expansion}
\end{equation}
and
\begin{equation}
\psi = \sum_{n=0}^\infty \alpha^n\psi_n(\xi, \eta, \tau,z),
\label{psi_expansion}
\end{equation}
where the $\zeta_n$'s and $\psi_n$'s are independent of $\alpha$.

We next insert the expansions (\ref{zeta_expansion}) and
(\ref{psi_expansion}) into Eqs. (\ref{Laplace}) - (\ref{eom})
and equate terms of the same order in $\alpha$.  The goal
will be to find a closed integral equation for $\zeta_0$.
We will need
to consider terms up to order $\alpha^3$.  To prepare
for this task, we note that to third order in $\alpha$, 
Eq. (\ref{topbc}) is

\begin{eqnarray}
\psi_z + \alpha\zeta\psi_{zz} &+& {1\over 2}\alpha^2\zeta^2\psi_{zzz}
+{1\over 6}\alpha^3\zeta^3\psi_{zzzz}\nonumber\\
&=&\alpha^2(\psi_\xi + \alpha\zeta\psi_{\xi z} - 1)\zeta_\xi 
+\alpha^2\zeta_\eta\psi_\eta 
+\alpha^3\zeta\zeta_\eta\psi_\eta.
\label{5}
\end{eqnarray}
This result holds for $z=1$.  To second order in $\alpha$, Eq. (\ref{eom})
reduces to
\begin{equation}
\alpha\zeta_\xi(\xi, \eta, \tau) + \sigma_q\alpha^2\zeta_\tau(\xi, \eta, \tau) 
= (\partial^2_\xi + \partial^2_\eta)\psi(\xi, \eta, 1+\alpha\zeta, \tau).
\label{6}
\end{equation}

We begin by working to zeroth order in $\alpha$.  To this order, 
Eq. (\ref{Laplace}) becomes $\psi_{0zz}=0$
for $0\le z\le 1$.  This implies that $\psi_{0z}$ does not depend on 
$z$.  Applying Eq. (\ref{bottombc}), we see that in fact $\psi_{0z}=0$
for all $z$, $\xi$, $\eta$ and $\tau$.  We conclude that $\psi_0$ depends only on
$\xi$, $\eta$ and $\tau$, and we write
\begin{equation}
\psi_0 = \theta_0(\xi, \eta, \tau).
\label{7} 
\end{equation}
The boundary condition at the free surface of the film gives no new information
to zeroth order in $\alpha$, as is readily seen from Eq. (\ref{5}).  

We now insert the expansions (\ref{zeta_expansion}) and (\ref{psi_expansion})
into Eq. (\ref{6}) and simplify using the fact that $\psi_{0z}=0$.  This gives
\begin{eqnarray}
\alpha\zeta_\xi + \sigma_q\alpha^2\zeta_\tau 
=\psi_{\xi\xi} + \psi_{\eta\eta} 
+\alpha^2(&\zeta_{0}&\psi_{1\xi\xi z} + 2 \zeta_{0\xi}\psi_{1\xi z} + \zeta_{0\xi\xi}\psi_{1z}
\nonumber\\
&+&\zeta_{0}\psi_{1\eta\eta z} + 2 \zeta_{0\eta}\psi_{1\eta z} + \zeta_{0\eta\eta}\psi_{1z} )
\label{8}
\end{eqnarray}
for $z=1$.  Eq. (\ref{8}) is valid to second order in $\alpha$.

Armed with this result, we return to the zeroth order analysis.  To that order,
Eq. (\ref{8}) reduces to 
\begin{equation}
\theta_{0\xi\xi} + \theta_{0\eta\eta}= 0.
\label{9}
\end{equation}
Now both $\theta_{0\xi}$ and $\theta_{0\eta}$
must vanish for $\vert\xi\vert\to\infty$.  The only
solution to Eq. (\ref{9}) that satisfies these boundary conditions is 
the trivial one: $\theta_0$ is a constant for all $\xi$ and $\eta$.  Because the
zero of the electrical potential is arbitrary, we may set this constant to zero.

To first order in $\alpha$, we find that $\psi_1$ is independent of $z$.
We shall write
\begin{equation}
\psi_1 = \theta_1(\xi, \eta, \tau).
\label{10}
\end{equation}
Eq. (\ref{5}) again yields no new information.  
The equation of motion (\ref{8}) is to first order
\begin{equation}
\theta_{1\xi\xi}+\theta_{1\eta\eta}=\zeta_{0\xi}\, .
\label{11}
\end{equation}

Our next task will be to write out the equations of motion to 
order $\alpha^2$.  Eqs. (\ref{Laplace}) and (\ref{11}) show that
\begin{equation}
\psi_{2zz} = -(\theta_{1\xi\xi}+\theta_{1\eta\eta})=-\zeta_{0\xi}
\end{equation} 
for $0\le z\le 1$.  Integrating this with respect to $z$
and using the boundary condition at 
the metal-insulator interface (\ref{bottombc}),
we have $\psi_{2z}=-\theta_{0\xi}z$.  Integrating once again, we obtain
\begin{equation}
\psi_2 = \theta_2 - {1\over 2}\zeta_{0\xi}z^2 ,
\label{12}
\end{equation}
where $\theta_2$ depends only on $\xi$, $\eta$ and $\tau$.  Once more,
Eq. (\ref{5}) tells us nothing new.  Eq. (\ref{8}) simplifies considerably
now that we know that $\psi_{1z}=0$.  Using Eq. (\ref{12}) as well, we obtain 
\begin{equation}
\sigma_q\zeta_{0\tau} + 
{1\over 2}(\zeta_{0\xi\xi\xi} + \zeta_{0\xi\eta\eta})
= \theta_{2\xi\xi} + \theta_{2\eta\eta} - \zeta_{1\xi}.
\label{13}
\end{equation}

Equating terms of order $\alpha^3$ in Eq. (\ref{Laplace}) yields
$\psi_{3zz}+\psi_{2\xi\xi}+\psi_{2\eta\eta} = 0$ for $0\le z\le 1$.  Integrating this twice
with respect to $z$ and applying Eqs. (\ref{bottombc}) and (\ref{12}),
we find that
\begin{equation}
\psi_3 = \theta_3 - {1\over 2}(\theta_{2\xi\xi} + \theta_{2\eta\eta}) z^2 
+ {1\over {24}}(\zeta_{\xi\xi\xi} + \zeta_{0\xi\eta\eta}) z^4 ,
\label{14}
\end{equation} 
where $0\le z\le 1$ and $\theta_3 = \theta_3(\xi, \eta, \tau)$.
We next equate terms of third order in Eq. (\ref{5}).  We obtain
\begin{equation}
\psi_{3z}+\zeta_0\psi_{2zz} = -\zeta_{1\xi} + \psi_{1\xi}\zeta_{0\xi} + 
\zeta_{0\eta}\psi_{1\eta} 
\label{15}
\end{equation}
for $z=1$.  This can be simplified using Eqs. (\ref{10}), (\ref{12}),
and (\ref{14}).  The result is 
\begin{equation}
\theta_{2\xi\xi} + \theta_{2\eta\eta} -\zeta_{1\xi} = 
{1\over 6}(\zeta_{0\xi\xi\xi} + \zeta_{0\xi\eta\eta}) -
\zeta_0\zeta_{0\xi} -\theta_{1\xi}\zeta_{0\xi}-\theta_{1\eta}\zeta_{0\eta}.
\label{16}
\end{equation}
Comparing this with Eq. (\ref{13}), we have
\begin{equation}
\sigma_q\zeta_{0\tau} + {1\over 3} (\zeta_{0\xi\xi} + \zeta_{0\eta\eta})_\xi
+ \zeta_0\zeta_{0\xi} = -(\theta_{1\xi}\zeta_{0\xi}+\theta_{1\eta}\zeta_{0\eta}).
\label{17}
\end{equation}
Since we have assumed that the disturbance is localized,
$\zeta_0$, $\theta_{1\xi}$ and $\theta_{1\eta}$ 
must all tend to zero as $\xi\to\pm\infty$.

Equations (\ref{11}) and (\ref{17}) are a closed system of equations that 
describe the dynamics of the film surface in the small amplitude, long wavelength
limit.  These equations represent a considerable simplification of the problem:
in the original formulation, the 3D Laplace equation for
$\Phi=\Phi(x,y,z,t)$ would have to be solved subject to boundary conditions at a moving 
boundary, a stationary plane boundary, 
and at infinity.  In the simplified description of the problem,
a Poisson equation for $\theta_1=\theta_1(\xi, \eta, \tau)$ 
[i.e., Eq. (\ref{11})] must be solved 
subject only to the boundary conditions at infinity.

If the film height is independent of $\eta$, Eq. (\ref{11}) reduces
to $\theta_{1\xi\xi}=\zeta_{0\xi}$.
Integrating this with respect to $\xi$ yields $\theta_{1\xi} =
\zeta_0 + F$, where $F$ is a function of $\tau$.  
Both $\theta_{1\xi}$ and $\zeta_0$ 
tend to zero as $\xi$ becomes large, and hence $F(\tau)=0$ for all $\tau$.
Eq. (\ref{17}) may therefore be rewritten as follows:
\begin{equation}
\sigma_q\zeta_{0\tau} + {1\over 3} \zeta_{0\xi\xi\xi} 
+ 2\zeta_0\zeta_{0\xi} = 0.
\label{18}
\end{equation}
Thus, when the film height is independent of $\eta$, the
equation of motion is the KdV equation, just as we found in Ref. \cite{Bradley}.

Before continuing, we shall simplify our notation somewhat.  Let 
$X=\sigma_q\xi$, $Y=\eta$, $T=\tau/3$, $u=6\zeta_0$, and
$\Psi=6\sigma_q\theta_1$.  Equations (\ref{11}) and (\ref{17}) become
\begin{equation}
u_T + (\partial_X^2 + \partial_Y^2)u_X + {1\over 2}uu_X
=-{1\over 2}(\Psi_X u_X + \Psi_Y u_Y)
\label{u_eqn}
\end{equation}
and
\begin{equation} \Psi_{XX}+\Psi_{YY}=u_X.
\label{Psi_eqn}
\end{equation}
The solution to Eq. (\ref{Psi_eqn}) that satisfies the requirement that
$\Psi_X\hat x+\Psi_Y\hat y$ tend to zero for $X\to\pm\infty$ is
\begin{equation}
\Psi(X,Y)={1\over{4\pi}}\int_{-\infty}^\infty dX' \int_{-\infty}^\infty dY'
\ln[(X-X')^2+(Y-Y')^2]u_X(X',Y').
\end{equation}
Insertion of this result into Eq. (\ref{u_eqn}) yields a closed integral equation
for $u$; finding such a relation was the goal of this section.  However, we will find it more 
convenient to work with Eqs. (\ref{u_eqn}) and (\ref{Psi_eqn}) rather than this unwieldy
integral equation.

SEM does not create or destroy atoms --- it simply moves them across the
metal surface.  The total mass, which is proportional to 
$\int_{-\infty}^\infty \int_{-\infty}^\infty h(x, y, t)dx dy$,
is therefore conserved.  This conservation law was not
lost in the asymptotic analysis:  
Eqs. (\ref{u_eqn}) and (\ref{Psi_eqn}) imply that 
$\int_{-\infty}^\infty \int_{-\infty}^\infty u(X, Y, T) dX dY$ 
is a constant. Indeed, using Eq. (\ref{Psi_eqn}), Eq. (\ref{u_eqn}) can be written
\begin{equation}
u_T + \vec\nabla\cdot\vec J = 0,
\label{u_cons}
\end{equation}
where $\vec\nabla \equiv \partial_X \hat X + \partial_Y \hat Y$ and
\begin{equation}
\vec J\equiv \vec\nabla u_X + {1\over 2}u \vec\nabla\Psi.
\label{J}
\end{equation}
Eq. (\ref{u_cons}) is the local statement of mass conservation and $\vec J$
is the current of the conserved quantity $u$.

Equations (\ref{u_eqn}) and (\ref{Psi_eqn}) have soliton solutions given by
\begin{equation}
u=12 a^2{\rm sech}^2[a(X-X_0-4a^2T)]
\label{19}
\end{equation}
and
\begin{equation}
\Psi=12a\tanh[a(X-X_0-4a^2T)],
\label{20}
\end{equation}
where $a$ and $X_0$ are constants.  In the next section, we will demonstrate
that these one-dimensional solitons are unstable against long-wavelength perturbations. 

If we set the right hand side of Eq. (\ref{u_eqn}) to zero, we obtain the ZK equation.
Thus, our problem differs from the ZK equation through the coupling to the field $\Psi$.
The linearized versions of the ZK equation and Eq. (\ref{u_eqn}) are the same.
Moreover, as we shall see, the 1D solitons are unstable in 
our problem, just as they are for the ZK equation \cite{LS,Infeld1,Infeld2,FI,AR}.  
However, there are important differences between the two problems: 
the growth rate of a perturbation of small transverse number $k$
will be shown to increase as $k^2$  
in our problem.  In contrast, this growth rate is proportional
to $\vert k\vert$ for small $k$ in the case of the ZK equation.

\section{Multiple scale perturbation analysis}

To prepare to investigate the stability of the soliton described by Eqs. (\ref{19})
and (\ref{20}), we shall transform to a coordinate system that moves along with the
soliton, and rescale so that the soliton amplitude is independent of $a$.
Specifically, we set 
$X-X_0=a^{-1}(\tilde X + 4\tilde T)$,
$Y=a^{-1}\tilde Y$,
$T=a^{-3}\tilde T$,
$u=a^2\tilde u$,
and $\Psi=a\tilde \Psi$ in Eqs. (\ref{u_eqn}) and (\ref{Psi_eqn}).  We shall again drop the
tildes, and, in addition, replace $X$, $Y$ and $T$ by $x$, $y$ and $t$.
This yields
\begin{equation}
u_t-4u_x+\nabla^2 u_x+{1\over 2}uu_x = -{1\over 2}(\Psi_x u_x+\Psi_y u_y)
\label{21}
\end{equation}
and
\begin{equation}
\nabla^2\Psi=u_x,
\label{22}
\end{equation}
where $\nabla^2\equiv \partial^2/\partial x^2 + \partial^2/\partial y^2$.
The unperturbed one-dimensional soliton is given by
\begin{equation}
u_0=12\,{\rm sech}^2 x
\label{23}
\end{equation}
and
\begin{equation}
\Psi_0=12\tanh x.
\label{24}
\end{equation}
Note that
\begin{equation}
u_{0xxx}-4 u_{0x}+u_0 u_{0x}=0;
\label{u0_identity}
\end{equation}
we will need this identity later.

We now apply a sinusoidal perturbation with wavevector
perpendicular to the direction of propagation: we put
\begin{equation} 
u=u_0 + \epsilon f(x)e^{iky+\gamma t}
\label{25}
\end{equation}
and
\begin{equation}
\Psi=\Psi_0+\epsilon\rho(x)e^{iky+\gamma t}
\label{26}
\end{equation}
and work to order $\epsilon$.  Eq. (\ref{21}) becomes
\begin{equation} 
{d\over{dx}}{\hat L}f-k^2f_x-{1\over 2} u_{0x}(f-\rho_x)=-\gamma f,
\label{f_eqn}
\end{equation}
where ${\hat L}\equiv d^2/dx^2+u_0-4$.  From Eq. (\ref{22}) we obtain
\begin{equation}
\rho_{xx}-k^2\rho=f_x.
\label{rho_eqn}
\end{equation}

We are interested in the effect of {\it localized} perturbations to
the 1D soliton.  Therefore, we shall seek solutions to Eqs. (\ref{f_eqn}) and (\ref{rho_eqn})
with the following properties: we require that
\begin{equation}
f\to 0 \,\,\, {\rm as} \,\, x\to\pm\infty
\label{f_bc}
\end{equation}
and
\begin{equation}
\rho_x\to 0 \,\,\, {\rm as} \,\, x\to\pm\infty .
\label{rho_bc}
\end{equation}                                                                                                  

We can obtain an approximate solution to Eqs. (\ref{f_eqn}) - (\ref{rho_bc}) 
if we restrict our attention to perturbations
with small transverse wavenumber $k$ and make a perturbation
expansion in powers of $k$ \cite{ZR}.  There is a technical difficulty, however.
In a standard perturbation analysis, we would set
\begin{equation}
f=f_0+kf_1+k^2f_2+\ldots,
\label{f_expansion}
\end{equation}
\begin{equation}
\rho=\rho_0+k\rho_1+k^2\rho_2+\ldots
\label{rho_expansion}
\end{equation}
and
\begin{equation}
\gamma=k\gamma_1+k^2\gamma_2+\ldots 
\label{gamma_expansion}
\end{equation}
(The zeroth order term in the expansion (\ref{gamma_expansion}) vanishes because
$\gamma=0$ for $k=0$ \cite{Jeffery,Benjamin}.)  We would then insert 
Eqs. (\ref{f_expansion}) - (\ref{gamma_expansion})
in Eqs. (\ref{f_eqn}) and (\ref{rho_eqn}) and equate the coefficients
of like powers of $k$.
Problems begin at the second order in the
perturbation expansion: $f_2$ cannot be made to vanish for both $x\to +\infty$
and $x\to -\infty$.  
The difficulties grow still worse after this point, since $f_3$ and higher order terms
diverge for large $x$.

To overcome these difficulties, we will employ a multiple scale perturbation
analysis \cite{Nayfeh}.  Our method is closely related to the multiple scale analysis 
developed by Allen and Rowlands for the ZK equation \cite{AR}.

The basic problem is that even though $k$ is small, $x$ can be large. In our 
multiple scale analysis, we develop
asymptotic expansions for $f$ and $\rho$.  If we keep terms of increasingly higher
order in these expansions, we obtain approximate expressions for $f$ and $\rho$
valid for larger and larger values of $x$.  To begin, we set $x_n\equiv k^n x$ for $n\ge 1$,
and consider $x$, $x_1$, $x_2$, and so forth to be independent variables.  Thus,
$f$ and $\rho$ are considered to be functions of $x$ and all of the $x_i$'s, and for 
small $k$ we have
\begin{equation}
f = f_0(x, x_1, x_2,\ldots) + k f_1(x, x_1, x_2,\ldots) + k^2 f_2(x, x_1, x_2,\ldots)+\ldots ,
\label{f_msexp}
\end{equation}
and
\begin{equation}
\rho = \rho_0(x, x_1, x_2,\ldots) + 
k \rho_1(x, x_1, x_2,\ldots) + k^2 \rho_2(x, x_1, x_2,\ldots)+\ldots .
\label{rho_msexp}
\end{equation}
Using the chain rule, we see that
\begin{equation}
{d\over{dx}} = {\partial\over{\partial x}} + k {\partial\over{\partial x_1}}
+ k^2 {\partial\over{\partial x_2}}+\ldots
\label{deriv_msexp}
\end{equation}
We insert Eqs. (\ref{f_msexp}) - 
(\ref{deriv_msexp}) and (\ref{gamma_expansion}) into Eqs. (\ref{f_eqn}) and (\ref{rho_eqn}) 
and equate coefficients of like powers of $k$.  To completely determine 
the $f_i$'s and $\rho_i$'s, additional constraints must be imposed.
We shall require that $f_m/f_{m-1}$ remain bounded as $x\to\infty$ for all
$m\ge 1$ \cite{Nayfeh}.  This ensures that $k^m f_m$ is small compared to
$k^{m-1} f_{m-1}$.  

Since $f$ and $\partial_x\rho$ tend to zero as $x\to\infty$, we demand that 
\begin{equation}
f_m\to 0 \quad {\rm as} \,\, x\to\infty
\label{f_m_bc}
\end{equation}
and
\begin{equation}
\partial_x\rho_m\to 0 \quad {\rm as} \,\, x\to\infty 
\label{rho_m_bc}
\end{equation}
for all $m\ge 0$.
Following Allen and Rowlands \cite{AR}, 
we shall also insist that
\begin{equation}
e^{2x} f_m\to 0 \quad {\rm as} \,\, x\to -\infty 
\label{f_m_bc_-}
\end{equation} 
for all $m\ge 0$.  The resulting expressions for
$f=f_0+kf_1+k^2f_2+\ldots$ and 
$\partial_x\rho=\partial_x\rho_{0}+k\partial_x\rho_{1}+k^2\partial_x\rho_{2}+\ldots$ 
will tend to zero as $x\to -\infty$, in accord with the
boundary conditions (\ref{f_bc}) and (\ref{rho_bc}).

If we retain $n$ terms in each of the asymptotic expansions (\ref{f_msexp}) and 
(\ref{rho_msexp}), we obtain expressions for $f$ and $\rho$ valid for $x$ 
as large as $k^{-n}$ \cite{Nayfeh}.  To show that the 1D solitons are unstable,
we will need to work to order $n=3$.  
For convenience,
we will write $f_{0x}\equiv\partial f_0/\partial x$,
$f_{01}\equiv\partial f_0/\partial x_1$,
$f_{0x1}\equiv\partial^2 f_0/\partial x\partial x_1$ and so on.

Before embarking on the multiple scale perturbation analysis in earnest,
we note that for large $\vert x\vert$, the problem simplifies considerably.  
Eq. (\ref{f_eqn}) reduces to
\begin{equation} 
f_{xxx}-(4+k^2)f_x+\gamma f=0
\label{f_simple}
\end{equation} 
in this limit.  Since $\gamma = O(k)$, 
the general solution to Eq. (\ref{f_simple}) is a linear combination
of $e^{p_1x}$, $e^{-p_2x}$ and $e^{p_3x}$, where
\begin{equation}
p_1=2-{1\over 8} \gamma +O(k^2),
\end{equation}
\begin{equation}
p_2=2+{1\over 8} \gamma +O(k^2), 
\end{equation}
and
\begin{equation}
p_3={1\over 4}\gamma +O(k^3).
\end{equation} 
Recall that $f$ must vanish for $x\to\pm\infty$.  Therefore, for large positive $x$,
\begin{equation}
f\cong A_+ e^{-p_2x} + B_+ e^{p_3x},
\label{f_+}
\end{equation}
while, for large negative $x$,
\begin{equation}
f\cong A_- e^{p_1 x} + B_- e^{p_3 x}.
\label{f_-}
\end{equation}
Note that if the real part of $p_3$ is positive, $B_+$ must vanish, and, similarly,
$B_-$ must be zero if the real part of $p_3$ is negative.  Eq. (\ref{f_-}) will later
provide a check of our analysis.

\subsection{Zeroth order}

To zeroth order, Eq. (\ref{f_eqn}) becomes
\begin{equation}
\partial_x L f_0 - {1\over 2} u_{0x}(f_0-\partial_x\rho_0) = 0, 
\label{f0_start}
\end{equation}
where 
\begin{equation} 
L\equiv {{\partial^2}\over{\partial x^2}} + u_0 -4.
\label{L_defn}
\end{equation}
[Note that ${\hat L}$ and $L$ differ: the partial
derivative with respect to $x$ appears in Eq. (\ref{L_defn}), 
whereas it is the total
derivative with respect to $x$ that appears in the definition of ${\hat L}$.]
Eq. (\ref{rho_eqn}) yields 
$\partial_x^2\rho_0 = \partial_x f_0$,
which may be integrated to give $\partial_x\rho_0 = f_0 + c_0$, where
$c_0 = c_0(x_1,x_2,\ldots)$ is independent of $x$.  Eqs. (\ref{f_m_bc}) and 
(\ref{rho_m_bc}) show that $c_0$ must vanish.  Hence
\begin{equation}
\rho_{0x}=f_0
\label{rho0_start}
\end{equation}
and Eq. (\ref{f0_start}) reduces to 
\begin{equation}
\partial_x L f_0 = 0.
\label{f0_new}
\end{equation}
This is a third order ordinary differential equation for $f_0$;
the general solution is
\begin{equation}
f_0 = h_0 \,{\rm sech}^2 x \tanh x + A_0 \chi + B_0 \cosh^2 x,
\end{equation}
where
\begin{equation}
\chi\equiv {1\over 4} (3x\,{\rm sech}^2 x \tanh x - 3\,{\rm sech}^2 x + 1).
\end{equation}
$h_0$, $A_0$ and $B_0$ are independent of $x$, but they 
may be functions of the $x_n$'s with $n\ge 1$.  Because $f_0$ must tend to
zero for $x\to +\infty$, the coefficients $A_0$ and $B_0$ must vanish.
We conclude that 
\begin{equation}
f_0 = h_0 F_0,
\label{f_0}
\end{equation}
where $F_0\equiv{\rm sech}^2 x\tanh x$.  Eq. (\ref{rho0_start}) now gives
$\rho_0 = -{1\over 2} h_0 {\rm sech}^2 x + c_0'$,
where $c_0'$ is independent of $x$.  Choosing the potential zero appropriately,
we can arrange to have $c_0'=0$, and then
\begin{equation}
\rho_0 = -{1\over 2} h_0 {\rm sech}^2 x .
\label{rho_0}
\end{equation}

\subsection{First order}

We next work to first order in $k$.  Using the identity $LF_0=0$,
Eq. (\ref{f_eqn}) yields 
\begin{equation}
\partial_x L f_1= - 2 f_{0xx1} + {1\over 2} u_{0x}(f_1-\rho_{1x}-\rho_{01})-\gamma_1 f_0 .
\label{f1_start}
\end{equation}
Eq. (\ref{rho_eqn}) becomes
$\rho_{1xx} + 2\rho_{0x1}=f_{1x}+f_{01}$.  Using Eqs. (\ref{f_0}) and (\ref{rho_0})
and integrating with respect to $x$, we find that
\begin{equation}
f_1-\rho_{1x}-\rho_{01}=c_1,
\label{rho1_start}
\end{equation}
where $c_1=c_1(x_1, x_2,\ldots)$ is independent of $x$.  This allows us to simplify
Eq. (\ref{f1_start}); we find that
\begin{equation}
\partial_xLf_1=-2h_{01}F_{0xx}-(\gamma_1 h_0+12c_1)F_0.
\label{f1_2}
\end{equation}
Eq. (\ref{f1_2}) has the solution
\begin{equation}
f_1=-h_{01}xF_0+{1\over{24}}\gamma_1h_0+{1\over 2}c_1+A_1\chi+B_1\cosh^2 x,
\label{f1_3}
\end{equation}
where $A_1=A_1(x_1,x_2,\ldots)$ and $B_1=B_1(x_1,x_2,\ldots)$ 
do not depend on $x$.\footnote{Strictly speaking, a term proportional to $F_0$ should appear
on the right hand side of this equation.  However, if this term is included,
the sole effect is to alter $f$ and $\rho$ by an irrelevant constant prefactor.
Therefore, without loss of generality, we may omit this term.}
Since $f_1$ must tend to zero as $x\to \infty$, the coefficient $B_1$ is zero.
Eq. (\ref{f1_3}) is therefore
\begin{equation}
f_1=\left({3\over 4}A_1 -h_{01}\right)xF_0 - {3\over 4}A_1\,{\rm sech}^2 x
+{1\over 4} A_1 + {1\over{24}}\gamma_1 h_0 +{1\over 2} c_1.
\label{f1_4}
\end{equation}
Now $f_1/f_0$ must remain bounded as $x\to\infty$; as a result
\begin{equation}
A_1={4\over 3}h_{01},
\label{A_1}
\end{equation}
\begin{equation}
c_1=-{1\over 2} A_1 - {1\over {12}}\gamma_1 h_0,
\label{c_1}
\end{equation}
and Eq. (\ref{f1_4}) reduces to 
\begin{equation}
f_1=-h_{01}\,{\rm sech}^2x.
\label{f1_5}  
\end{equation}
Using Eq. (\ref{A_1})
to eliminate $A_1$ from Eq. (\ref{c_1}), we find that
\begin{equation}
h_{01}+{1\over 8}\gamma_1h_0+{3\over 2}c_1 = 0.
\label{h01_start}
\end{equation}

We now turn our attention to Eq. (\ref{rho1_start}).  Using Eqs. (\ref{rho_0}) and (\ref{f1_5}),
we obtain $\rho_{1x}=-{1\over 2}h_{01}{\rm sech}^2 x - c_1$.
The boundary condition (\ref{rho_m_bc}) is satisfied only if $c_1=0$; Eq. (\ref{h01_start})
therefore becomes
\begin{equation}
h_{01}= -{1\over 8}\gamma_1h_0.
\label{h01}
\end{equation}
We conclude that 
\begin{equation}
f_1={1\over 8}\gamma_1h_0\,{\rm sech}^2x
\label{f1}  
\end{equation}
and
\begin{equation}
\rho_{1x}={1\over {16}}\gamma_1 h_0\,{\rm sech}^2x.
\end{equation}
We integrate the latter equation with respect to $x$ and choose the
potential zero so that $\rho_1\to 0$ as $x\to\infty$.  This yields
\begin{equation}
\rho_1={1\over{16}}\gamma_1 h_0(\tanh x-1).
\label{rho1}
\end{equation}

\subsection{Second Order}

Equating coefficients of terms of second order in $k$ in Eq. (\ref{f_eqn})
and using the identity $LF_0=0$, we have
\begin{eqnarray}
\partial_xL f_2 = &-&2 f_{1xx1}-3 f_{0x11} -2 f_{0xx2}-L f_{11} + f_{0x}\nonumber\\
			&+&{1\over 2}u_{0x}(f_2-\rho_{2x}-\rho_{11}-\rho_{02})
			-\gamma_1 f_1 -\gamma_2 f_0 .
\label{f2_start}
\end{eqnarray}
Eq. (\ref{rho_eqn}) yields
\begin{equation}
f_{2x}-\rho_{2xx}-\rho_{1x1}-\rho_{0x2}={1\over 2}h_0{\rm sech}^2 x;
\label{rho2_start}
\end{equation}
this result has been simplified using the fact that $c_1=0$ and
Eqs. (\ref{rho0_start}), (\ref{rho_0}), and (\ref{rho1_start}).
Integrating Eq. (\ref{rho2_start}) with respect to $x$, we obtain
\begin{equation}
f_2-\rho_{2x}-\rho_{11}-\rho_{02}={1\over 2}h_0\tanh x + c_2,
\label{rho2_2}
\end{equation}
where $c_2=c_2(x_1,x_2,\ldots)$.
We now insert Eqs. (\ref{f_0}), (\ref{rho_0}), (\ref{h01}), (\ref{f1}),
(\ref{rho1}) and (\ref{rho2_2})
into Eq. (\ref{f2_start}).  This gives
\begin{eqnarray}
\partial_xL f_2 &=& \left(1-{7\over{64}}\gamma_1^2\right)h_0F_{0x}-2h_{02}F_{0xx}
-(\gamma_2h_0+12c_2)F_0\nonumber\\
& &-{1\over{96}}\gamma_1^2h_0u_0
+{1\over{768}}\gamma_1^2h_0Lu_0 - 6h_0F_0\tanh x.
\label{f2_2}
\end{eqnarray}

The last two terms on the right hand side of Eq. (\ref{f2_2}) must be rewritten
before we can proceed further.  To begin, note that Eq. (\ref{u0_identity}) can be
integrated to yield
\begin{equation}
u_0^2=8u_0+48F_{0x}.
\label{u0_id2}
\end{equation}
Using this identity, it can readily be shown that
\begin{equation}
Lu_0=24F_{0x}+4u_0
\end{equation}
and 
\begin{equation}
36 F_0\tanh x = u_0-12F_{0x}.
\end{equation}
Eq. (\ref{f2_2}) now becomes
\begin{eqnarray}
\partial_xL f_2 &=& \left(3-{5\over{64}}\gamma_1^2\right)h_0F_{0x}-2h_{02}F_{0xx}
-{1\over{6}}\left(1+{1\over{32}}\gamma_1^2\right)h_0u_0\nonumber\\
& &-(\gamma_2h_0+12c_2)F_0.
\label{f2_3}
\end{eqnarray}
Eq. (\ref{f2_3}) has the solution
\begin{eqnarray}
f_2&=&-{1\over 6}\left ({2\over 5}+{3\over{64}}\gamma_1^2\right )h_0\cosh x\sinh x
	+{1\over{10}}\left(3-{5\over{64}}\gamma_1^2\right)h_0\tanh x
	\nonumber\\
     & &+{1\over{24}}(\gamma_2h_0+12c_2)-h_{02}xF_0
	+A_2h_0\chi 
	+B_2\cosh^2 x,
\label{f2_4}
\end{eqnarray}
where $A_2=A_2(x_1,x_2,\ldots)$ and $B_2=B_2(x_1,x_2,\ldots)$ 
do not depend on $x$.\footnote{Footnote 1 again applies.} 

Because $f_2$ must satisfy the boundary conditions (\ref{f_m_bc}) and (\ref{f_m_bc_-}), 
the coefficients of 
$\cosh x\sinh x$ and $\cosh^2 x$ in Eq. (\ref{f2_4}) must vanish:
\begin{equation}
\gamma_1^2=-{{128}\over {15}}
\label{gamma_1}
\end{equation}
and
\begin{equation}
B_2=0.
\label{B_2}
\end{equation}
Eq. (\ref{f2_4}) now reduces to 
\begin{eqnarray}
f_2 &=& {{11}\over{30}}h_0\tanh x
     +{1\over{24}}(\gamma_2h_0+12c_2+6 A_2 h_0)+
      \left({3\over 4}A_2 h_0- h_{02}\right)xF_0\nonumber\\
    & &  - {3\over 4}A_2 h_0\,{\rm sech}^2 x.
\label{f2_5}
\end{eqnarray}
$f_2$ must tend to zero as $x\to\infty$, and therefore
\begin{equation}
\left({{44}\over 5} +\gamma_2 + 6A_2\right)h_0+12 c_2=0.
\label{limit1}
\end{equation}
Since $f_2/f_1$ must remain bounded as $x\to\infty$, we must also have
\begin{equation}
h_{02}={3\over 4}A_2h_0.
\label{limit2}
\end{equation}

All of the terms on the left hand side of Eq. (\ref{rho2_2}) tend to zero 
for $x\to\infty$, and so
\begin{equation}
c_2=-{1\over 2}h_0.
\label{c_2}
\end{equation}
Eliminating $c_2$ from Eq. (\ref{limit1}) using this result and $A_2$ using
Eq. (\ref{limit2}), we obtain
\begin{equation}
h_{02}=-{1\over 4}\left({7\over 5}+{1\over 2}\gamma_2\right)h_0.
\label{h_02}
\end{equation}
Eq. (\ref{f2_5}) can now be simplified using Eqs. (\ref{limit1}), (\ref{limit2}), and (\ref{h_02}).
We find that
\begin{equation}
f_2={1\over 4}\left({7\over 5}+{1\over 2}\gamma_2\right)h_0\,{\rm sech}^2 x 
+ {{11}\over{30}}h_0(\tanh x -1).
\label{f_2}
\end{equation}
Notice that although $e^{2x}f_2$ tends to zero as $x\to -\infty$, $f_2$ itself 
does not vanish in that limit.

We can now determine $\rho_{2x}$ using 
Eqs. (\ref{rho_0}), (\ref{h01}), (\ref{rho1}), (\ref{c_2}), 
(\ref{h_02}), and (\ref{f_2}) in Eq. (\ref{rho2_2}).
The result is
\begin{equation}
\rho_{2x}={1\over 8}\left({7\over 5}+{1\over 2}\gamma_2\right)h_0\,{\rm sech}^2 x
-{1\over 5}h_0(\tanh x -1).
\label{rho_2x}
\end{equation}
Integrating this and choosing the potential zero to be at $x=+\infty$, we have
\begin{equation}
\rho_{2}={1\over 8}\left({7\over 5}+{1\over 2}\gamma_2\right)h_0(\tanh x-1)
-{1\over 5}h_0(\ln\cosh x -x+\ln 2).
\label{rho_2}
\end{equation}

\subsection{Third Order}

Using the identity $LF_0=0$ once again and equating coefficients of terms of 
order $k^3$ in Eq. (\ref{f_eqn}), we have
\begin{eqnarray}
\partial_x L f_3 & &+ \partial_{x_1} L f_2 + \partial_{x_2} L f_1
+2 f_{2xx1} + 3 f_{1x11} + 2 f_{1xx2} + 2 f_{0xx3} + 6 f_{0x12} \nonumber\\  
& &+ f_{0111} -f_{1x}-f_{01}+12 F_0(f_3-\rho_{3x}-\rho_{21}-\rho_{12}-\rho_{03})\nonumber\\
& &= -(\gamma_1 f_2 + \gamma_2 f_1 + \gamma_3 f_0).
\label{f3_1}
\end{eqnarray}
From Eq. (\ref{rho_eqn}), we obtain
\begin{eqnarray}
\rho_{3xx} + 2\rho_{2x1} &+& 2\rho_{1x2} + \rho_{111} 
+ 2\rho_{0x3} + 2\rho_{012} - \rho_{1}\nonumber\\ 
&=& f_{03} + f_{12} + f_{21} + f_{3x}.
\label{rho3_1}
\end{eqnarray}
This result can be simplified using Eqs. (\ref{rho0_start}), 
(\ref{rho1_start}), and (\ref{rho2_2}), and the values of $c_1$ and $c_2$.  This gives
\begin{equation}
f_{3x}-\rho_{3xx} -\rho_{2x1}-\rho_{1x2}-\rho_{0x3}=0,
\end{equation}
and, integrating, we obtain $f_{3}-\rho_{3x} -\rho_{21}-\rho_{12}-\rho_{03}= c_3$,
where $c_3=c_3(x_1,x_2,\ldots)$.  Now $\rho_0$, $\rho_1$, and $\rho_2$
tend to zero as $x\to +\infty$, and $f_3$ and $\rho_{3x}$ must vanish
in this limit as well.  Thus, $c_3=0$ and
\begin{equation}
f_{3}-\rho_{3x} -\rho_{21}-\rho_{12}-\rho_{03}= 0.
\label{rho3_start}
\end{equation}

We now return to Eq. (\ref{f3_1}).  Using Eq. (\ref{rho3_start}) and our 
explicit expressions for $\gamma_1^2$, $h_{01}$,
$h_{02}$, $f_0$, $f_1$, and $f_2$, we obtain 
\begin{eqnarray}
-\partial_x Lf_3=& &{{11}\over{20}} \gamma_1 h_0(\tanh x-1)
+{1\over{48}}\gamma_1\left({{11}\over 5} 
+{1\over 2}\gamma_2\right)h_0u_0\nonumber\\
& &+\left({{13}\over{60}}\gamma_1+\gamma_3\right)h_0F_0
+{5\over{16}}\left({7\over 5}+{1\over 2}\gamma_2\right)\gamma_1h_0F_{0x}\nonumber\\
& &+2h_{03}F_{0xx}.
\label{f3_2}
\end{eqnarray}
The solution to this equation is
\begin{eqnarray}
-f_3 =& & {1\over 4}\left(\gamma_2-{8\over{15}}\right)\gamma_1h_0\cosh x\sinh x\nonumber\\
& &+{{11}\over{80}}\gamma_1 h_0\Bigg [
{1\over 4}u_0(\ln\cosh x-x-1)\nonumber\\
& &+3 F_0\left({1\over 2}x^2-\int_0^x\ln\cosh x'\, dx'\right )
+x-\ln\cosh x\Bigg ]\nonumber\\
& &+{1\over{24}}\left({{251}\over {60}}\gamma_1-\gamma_3\right)
+{1\over{16}}\left({1\over 4}\gamma_2-{{13}\over 5}\right)\gamma_1 h_0\tanh x\nonumber\\
& &+h_{03}xF_0
+A_3h_0\chi+B_3\cosh^2 x,
\label{f3_3}
\end{eqnarray}
where $A_3$ and $B_3$ are independent of $x$.\footnote{Footnote 1 again applies.} 
Applying the boundary conditions (\ref{f_m_bc}) and (\ref{f_m_bc_-}) with $m=3$, 
we see that $B_3$ vanishes and
\begin{equation}
\gamma_2={8\over{15}}.
\label{gamma_2}
\end{equation}

Eq. (\ref{gamma_2}) is the principal result of this section.  We have now shown
that the one-dimensional solitons are unstable.

We must still show that our approximate solution can be given the
correct asymptotic behavior.  We require that $f_3\to 0$ as
$x\to +\infty$ and that $f_3/f_2$ remain bounded in this limit.
This yields
\begin{eqnarray}
f_3 =
& &{{11}\over{80}}\gamma_1 h_0\left[
-{1\over 4}u_0\ln(1+e^{-2x})
+3 F_0\int_0^x\ln(1+e^{-2x'})\, dx'
+\ln(1+e^{-2x})\right ]\nonumber\\
& &+{{37}\over {240}}\gamma_1 h_0(\tanh x-1)
+{1\over {96}}\left(\gamma_3+{{169}\over{60}}\gamma_1\right)h_0u_0.
\label{f3_4}
\end{eqnarray}
If $x$ is large in magnitude and negative,
\begin{equation} 
{{f_3}\over {h_0}}= -{{11}\over {40}}\gamma_1 x +O(1).
\label{f3_5}
\end{equation}
$e^{2x}f_{3}$ vanishes for $x\to -\infty$, but $f_{3}$ does not.

Suppose that $x$ is large and negative, $k$ is small, and $kx$ is of order 1.
Dropping terms of order $e^{2x}$, we have $f_0/h_0=f_1/h_0=0$ and 
$f_2/h_0=-11/15$.  Eq. (\ref{h01}) shows that
\begin{equation}
h_0=\exp\left(-{1\over 8}\gamma_1 x_1\right)h_1,
\label{h0}
\end{equation}
where $h_1 = h_1(x_2,x_3,\ldots)$.  In the limit we are presently considering,
$h_1$ is approximately constant.  Using these results and Eq. (\ref{f3_5}),
we obtain 
\begin{eqnarray}
f&=&f_0+kf_1+k^2f_2+k^3f_3+\ldots\nonumber\\
 &\cong&-{{11}\over{15}}h_1k^2\exp\left(-{1\over 8}k\gamma_1 x\right)
\left(1+{3\over 8}k\gamma_1 x\right)+k^4f_4+\ldots
\end{eqnarray}
Retaining terms up to third order in $k$, this may be written
\begin{equation}
f\cong -{{11}\over{15}}h_1k^2\exp\left({1\over 4}\gamma x\right).
\end{equation}
This is consistent with Eq. (\ref{f_-}), and so $f$ does indeed have
the correct asymptotic behavior in the $x\to -\infty$ limit.

\section{Conclusions}

In this paper, we found the equations of motion for small amplitude, long waves
on the surface of a current-carrying metal film.  In dimensionless units, the equations are
\begin{equation}
u_T + (\partial_X^2 + \partial_Y^2)u_X + {1\over 2}uu_X
=-{1\over 2}(\Psi_X u_X + \Psi_Y u_Y)
\label{u_eqn2}
\end{equation}
and
\begin{equation} 
\Psi_{XX}+\Psi_{YY}=u_X.
\label{Psi_eqn2}
\end{equation}
These equations represent a generalization of the KdV equation to two dimensions, 
and seem not to have been studied previously.

1D solitons traveling over the surface of a current-carrying
metal thin film were shown to be unstable.  
The growth rate for perturbations of small transverse 
wavevector $k$ is
\begin{equation}
{\rm Re}\, \gamma = {8\over{15}}k^2+O(k^3).
\end{equation}
For the ZK \cite{LS,Infeld1,Infeld2,FI,AR}
and KPII \cite{KP} equations, the growth rate ${\rm Re}\, \gamma$ increases
much faster than this as $k$ is increased: 
it is proportional to $\vert k\vert$ for small $k$.  

The linear stability analysis carried out in this paper only applies shortly 
after the 1D soliton is perturbed.  What is the nature of the
subsequent time evolution?  
For the ZK equation, it is known that if a 1D soliton is perturbed, it will break up into
a series of \lq\lq cylindrical" solitons \cite{Frycz}.  These solitons are propagating humps,
and are stable.  Numerical studies of Eqs. (\ref{u_eqn2}) and (\ref{Psi_eqn2}) are
currently underway, and will reveal whether a
1D soliton propagating over a current-carrying metal surface breaks up in a similar fashion
after it has been perturbed.

{\bf Acknowledgements}

I would like to thank Gerhard Dangelmayr, Marty Gelfand 
and especially Harvey Segur for useful discussions.

\end{document}